\documentclass[]{article}
\usepackage{latexsym,amssymb,amsmath,amsfonts}
\usepackage{pictex}

\newcommand{\nt}{\not \negthinspace}

\begin{document}

\vspace{10mm}

\centerline{\Large\bf Antisymmetric tensor contribution to the muon $g-2$}

\vspace{5mm}

\centerline{Terry~Pilling\footnote{terry@offshell.phys.ndsu.nodak.edu}
and Richard~T.~Hammond\footnote{rich.hammond@ndsu.nodak.edu} 
}
\centerline{{\it Department of Physics, North Dakota State University}}
\centerline{{\it Fargo, ND, 58105-5566}}
\bigskip
\centerline{P.F.~Kelly\footnote{pfkelly@avemaria.edu} }
\centerline{{\it Department of Physics, Ave Maria College}}
\centerline{{\it Ypsilanti, MI, 48197}}

\vspace{10mm}

\begin{abstract}

We investigate the Kalb-Ramond antisymmetric tensor field 
as a solution to the muon $g-2$ problem. 
In particular we calculate the lowest-order Kalb-Ramond 
contribution to the muon anomalous magnetic 
moment and find that we can fit the new experimental value for the 
anomaly by adjusting the coupling without affecting 
the electron anomalous magnetic moment results.

\end{abstract}

\section{Introduction}

Particle magnetic moments have provided a valuable test of
QED and the Standard Model. 
In the magnetic moments of the electron
and the muon we have the important situation that both
the experimental measurements and the standard model
predictions are extremely precise. 
Additional importance of the
muon stems from the fact that it has a larger rest mass
than the electron and hence is more sensitive to massive 
virtual particles and to new physics.
See \cite{hughes1999} for a review of the electron 
and the muon anomalous magnetic moments.

It was recently found that the complete standard 
model prediction for the muon magnetic moment differs 
from the experimentally determined value \cite{brown0102017,hertzog0202024} 
by 1.6 standard deviations \footnote{The discrepancy originally stated in 
\cite{brown0102017} and discussed in 
\cite{miller0111036,onderwater0105036} turned out to be 
incorrect due to a sign error found in the part of the standard
model calculation dealing with the pion pole contribution. 
We refer the interested reader to 
\cite{hertzog0202024,hayakawa0112102,blokland0112117,
knecht0111058,bijnens0112255} and references therein for the details.}. 
For a nice summary of the experimental and theoretical analyses,
as well as history and prospects, see \cite{miller0111036}.
The measured value is expressed as 
$a(\mbox{exp}) = 11659203 \pm 15 \times 10^{-10}$
(assuming CPT invariance) where $a = (g-2)/2$.
The standard model prediction is 
$a(\mbox{SM}) = 11659176.7 \pm 6.7 \times 10^{-10}$.

There are many possible extensions
of the standard model which may contribute to the anomalous
magnetic moment \cite{czarnecki,einhorn}. 
For example, the minimal supersymmetric extension of the 
standard model (MSSM) 
\cite{grifols1982,feng,czarnecki,everett0102145,ellis,ferrer}, 
higher dimensional Kaluza-Klein theories \cite{calmet}, 
as well as theories based on heretofore 
undiscovered interactions \cite{einhorn}, are all of this type.

In the present paper we investigate a new interaction involving an
antisymmetric tensor field which we present as a potential 
solution to the $1.6\sigma$ discrepancy. 

\section{The Antisymmetric Tensor Interaction}  
\label{rules}

\begin{figure}
\begin{center}
\input{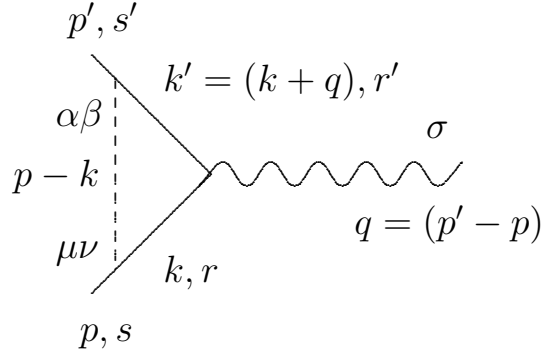}
\end{center}
\caption{The Feynman diagram for the antisymmetric tensor contribution 
to the muon anomalous magnetic moment. 
We have written the photon polarization index $\sigma$, 
the antisymmetric tensor indices $\mu$, $\nu$, $\alpha$, $\beta$, and the 
spinor indices $s$, $s'$, $r$, $r'$. The initial and final muon 
momenta are $p$ and $p'$, the momentum of the antisymmetric
tensor field is $p-k$ and the momentum transfered to the photon 
is $q$.
}
\label{moment}
\end{figure}

The Kalb-Ramond antisymmetric tensor field \cite{kalb,townsend} 
is encountered in string theory \cite{pilling} and also arises in 
some theories of gravity with torsion 
\cite{hammond2002}.
The contribution of the antisymmetric tensor to the anomalous magnetic
moment occurs at one-loop order due to the Kalb-Ramond--fermion vertex.
The Feynman diagram for this process is shown in Fig.~\ref{moment}.

Consider the following lagrangian \cite{pilling,dass0107006,miyamoto}:
\begin{equation}
\label{actionkalb}
\mathcal{L} = -\frac{1}{4} F_{\mu \nu} F^{\mu \nu} 
- \frac{1}{12} H_{\mu \nu \lambda} H^{\mu \nu \lambda} 
+ \overline{\psi} \left( i \nt{\partial} 
- e \nt{A} - m - \frac{g}{M} \sigma_{\mu \nu \lambda} H^{\mu \nu \lambda} 
\right) \psi.
\end{equation}
The factor of $M$ denotes
a mass scale to make the coupling constant, $g$, dimensionless.
We assume that the antisymmetric torsion, or Kalb-Ramond antisymmetric
tensor field, $H^{\mu \nu \lambda}$, can be derived from a potential
(i.e. $H = dB$ is exact)
\begin{equation}
\label{defs1}
H_{\mu \nu \lambda} = \partial_\mu B_{\nu \lambda} + \partial_\nu 
B_{\lambda \mu} + \partial_\lambda B_{\mu \nu}, 
\end{equation}
with $B_{\nu \lambda} = - B_{\lambda \nu}$, and
\begin{equation}
\sigma_{\alpha \beta \gamma} = i \epsilon_{\alpha \beta \gamma \mu} 
\gamma_5 \gamma^\mu. 
\end{equation}

To find the propagator for the antisymmetric tensor
field requires adding a gauge fixing term to the free 
lagrangian density
\begin{equation}
\begin{split}
\mathcal{L}_0 + \mathcal{L}_{\sf{gf}} &= 
- \frac{1}{12} H_{\mu \nu \lambda} H^{\mu \nu \lambda} 
- \frac{1}{4 \alpha} 
\left( \partial_\mu B^{\mu \nu} \partial^\sigma B_{\sigma \nu} 
+ \partial_\mu B^{\nu \mu} \partial^\sigma B_{\nu \sigma} \right) \\
&= \frac{1}{4} 
B^{\alpha \beta} \Bigl[ g_{\alpha \nu} g_{\beta \lambda} \Box 
+ g_{\beta \nu} \partial_\lambda \partial_\alpha 
+  g_{\alpha \lambda} \partial_\nu \partial_\beta 
+ \frac{1}{\alpha} 
\left( g_{\beta \lambda} \partial_\alpha \partial_\nu 
+ g_{\alpha \nu} \partial_\beta \partial_\lambda \right) \Bigr] 
B^{\nu \lambda},
\end{split}
\end{equation}
where $\alpha$ is a gauge fixing parameter.
We use the antisymmetry of $B_{\mu \nu}$
and substitute $i \partial_\mu \rightarrow k_\mu$ to arrive at the 
momentum space expression
\begin{equation}
\label{quadratic}
\mathcal{L}_0 + \mathcal{L}_{\sf{gf}} = 
- \frac{1}{4} B^{\alpha \beta} 
\Bigl[ g_{\alpha \nu} g_{\beta \lambda} k^2 
\quad + \left(\frac{1}{\alpha} - 1 \right) 
\left( g_{\beta \lambda} k_\alpha k_\nu 
+ g_{\alpha \nu} k_\beta k_\lambda \right) \Bigr] 
B^{\nu \lambda}.
\end{equation}

Our gauge invariant propagator for the antisymmetric tensor field is
the inverse of the quantity in brackets directly above and reads:
\begin{equation}
G^{\nu \alpha \lambda \beta} = \frac{1}{k^2} 
\Bigl[ g^{\nu \beta} g^{\lambda \alpha} - g^{\nu \alpha} g^{\lambda \beta} 
+ \frac{(1-\alpha)}{k^2} 
\left(g^{\nu \alpha} k^\lambda k^\beta 
+ g^{\lambda \beta} k^\nu k^\alpha 
- g^{\nu \beta} k^\lambda k^\alpha - g^{\lambda \alpha} k^\nu k^\beta 
\right) \Bigr] \ .
\end{equation}
With the gauge choice $\alpha = 1$ this reduces to
\begin{equation}
\label{propagator}
G^{\nu \alpha \lambda \beta}= \frac{1}{k^2} 
\left( g^{\nu \beta} g^{\lambda \alpha} 
- g^{\nu \alpha} g^{\lambda \beta} \right).
\end{equation}

The vertex rule for the coupling of the
antisymmetric tensor field to a spin-$\frac{1}{2}$ Dirac field 
is extracted from the interaction part of the lagrangian
\begin{equation}
\mathcal{L}_{\sf{int}} = -\frac{g}{M} \overline{\psi} \;
\sigma_{\mu \nu \lambda} H^{\mu \nu \lambda} \; \psi.
\end{equation}
Substituting the definitions (\ref{defs1}) into this gives
\begin{equation}
\begin{split}
\mathcal{L}_{\sf{int}} &= 
-\frac{g}{M} \overline{\psi} 
\left[ \left( i \epsilon_{\mu \nu \lambda \sigma} \gamma_5 \gamma^\sigma \right)
\left( \partial^\mu B^{\nu \lambda} + \partial^\nu B^{\lambda \mu} 
+ \partial^\lambda B^{\mu \nu} \right) \right] \psi \\
&\equiv -\frac{g}{M} \overline{\psi} \; \Lambda^{\alpha \beta} \; \psi 
\; B_{\alpha \beta}.
\end{split}
\end{equation}
Our vertex rule in momentum space is then seen to be
\begin{equation}
\label{vertexrule}
-\frac{g}{M} \Lambda^{\alpha \beta} = -\frac{g}{M} 
\epsilon_{\mu \nu \lambda \sigma} \gamma_5 \gamma^\sigma 
\left( g^{\nu \alpha} g^{\lambda \beta} k^\mu 
+ g^{\lambda \alpha} g^{\mu \beta} k^\nu 
+ g^{\mu \alpha} g^{\nu \beta} k^\lambda \right),
\end{equation}
where the momentum $k$ is that carried by the Kalb-Ramond field
$B_{\mu \nu}$.

\section{Fermion Anomalous Magnetic Moment}

The corrections to the magnetic moment are due to exchanges of 
virtual particles which affect the measured value of the 
photon-fermion coupling.
Here we specialize to the case of muons.

The QED prediction is (to 5 loops) 
\cite{brown0102017,hertzog0202024,kinoshita2}:
\[
a(\mbox{QED}) = 116584705.7(2.9) \times 10^{-11}
\]
which differs in the 5th significant figure from the experimental value. 
The electroweak corrections to the vertex amount to
(to 2 loops) \cite{brown0102017,czarnecki1,czarnecki2}:
\[
a(\mbox{EW}) = 152(4) \times 10^{-11}
\]

There is yet another standard model correction.
It is when the first muon emits a photon which is energetic
enough to produce 
quark anti-quark pairs which then annihilate
into another photon absorbed by the final state muon.
This hadronic correction to the moment is
computed two separate ways -- with results which are in agreement --
(to 3 loops) \cite{brown0102017,kinoshita1,erler}:
\begin{equation*}
\begin{split}
a(\mbox{hadron1}) &= 6739(67) \times 10^{-11}\\
\mbox{or } \qquad a(\mbox{hadron2}) &= 6803(114) \times 10^{-11}.
\end{split}
\end{equation*}
We must note that this is very difficult to calculate and that
it contributes the greatest amount of uncertainty in the total
standard model prediction.

So the complete standard model prediction is then:

\begin{equation}
\begin{split}
a(\mbox{SM}) = a(\mbox{QED}) + a(\mbox{EW}) + a(\mbox{hadron1}) 
&= 116591597(67) \times 10^{-11} \\
\mbox{ or (with a(\mbox{hadron2}))} &= 116591660(114) \times 10^{-11}
\end{split}
\end{equation}
We now see that this differs from experiment in the 6th
digit, which is much better than the pure QED calculation. 

The current standard model calculation differs from the latest 
experimental results and is outside of the experimental error bars. 
This motivates a search for sources beyond the standard model that
may help to resolve the discrepancy.

\section{Torsion contribution to the magnetic moment}

We can use the Feynman rules derived in Section~\ref{rules} to
find the contribution of the antisymmetric tensor field to fermion
anomalous magnetic moments. This may place a useful bound on the 
fermion-antisymmetric tensor coupling.

The vertex correction, shown in Fig.~\ref{moment}, is defined by
$\Gamma^\mu = \gamma^\mu + \delta \Gamma^\mu$, where
\begin{equation}
\delta \Gamma^\sigma \left(p', p \right) = \int 
\frac{d^4 k}{(2 \pi)^4} \Biggl\{ 
\frac{g^2}{m^2} \Lambda^{\alpha \beta}_{s' r'}(p-k) 
\left[ i G_{\alpha \beta \mu \nu}(p-k) \right] 
\Lambda^{\mu \nu}_{r s}(p-k) 
\left[ i S_F(k)_{l r} \right] 
\left( - \gamma^{\sigma}_{l' l} \right)
\left[ i S_F(k')_{r' l'} \right] \Biggr\}.
\end{equation}
Inserting the expressions for the propagator (\ref{propagator}) 
and vertex rule (\ref{vertexrule}) and using the standard identities
\begin{equation}
\begin{split}
\epsilon_{\mu \nu \alpha \beta} \epsilon^{\alpha \beta \lambda \sigma} &=
-4 \delta_{[\mu}^{\lambda} \delta_{\nu]}^{\sigma} \\
\gamma^\rho \nt{k}' \gamma^\sigma \nt{k} \gamma_\rho &= 
-2 \nt{k} \gamma^\sigma \nt{k}' + \epsilon \nt{k}' \gamma^\sigma \nt{k} \\
\gamma^\rho \gamma^\sigma \nt{k} \gamma_\rho &= 
4 k^\sigma - \epsilon \gamma^\sigma \nt{k} \\
\gamma^\rho \gamma^\sigma \gamma_\rho &= 
\left( \epsilon - 2 \right) \gamma^\sigma
\end{split}
\end{equation}
our vertex correction becomes
\begin{equation}
\delta \Gamma^\sigma \left(p', p \right) = 
\frac{i 36 g^2}{M^2} \int \frac{d^4 k}{(2 \pi)^4} \Biggl\{
\frac{A^\sigma}{\left[ (k')^2-m^2 \right] \left[ k^2-m^2 \right]} 
- \frac{B^\sigma}{\left[ k-p \right]^2 \left[ (k')^2-m^2 \right]}
\Biggr\},
\end{equation}
where 
\begin{equation}
\begin{split}
A^\sigma &= \left(8mk^\sigma + 4mq^\sigma -2 \nt{k} \gamma^\sigma
\nt{k} - 2 \nt{k} \gamma^\sigma \nt{q} - 2m^2 \gamma^\sigma \right) 
+ \epsilon \left(\nt{k} \gamma^\sigma \nt{k} 
+ \nt{q} \gamma^\sigma \nt{k} - 2m k^\sigma - m \nt{q} \gamma^\sigma 
+ m^2 \gamma^\sigma \right), \\
B^\sigma &= 2 p\cdot k \gamma^\sigma + 2 p \cdot q \gamma^\sigma
-2 p^\sigma \nt{k} - k^2 \gamma^\sigma - \nt{k} \nt{q} \gamma^\sigma 
+ 2 m p^\sigma - m^2 \gamma^\sigma.
\end{split}
\end{equation}

Power counting indicates a quadratic divergence in
need of regularization.
Feynman parametrization to combine denominator factors gives
\begin{equation}
\delta \Gamma^\sigma = 
\frac{i 36 g^2}{M^2} \int \frac{d^4 k}{(2 \pi)^4} 
\int_0^1 dx \;dy\; \delta(x+y-1) 
\Biggl\{
\frac{N_1^\sigma + \epsilon N_2^\sigma}{\left[k^2 - \Delta_1 \right]^2} 
- \frac{N_3^\sigma}{\left[k^2 - \Delta_2 \right]^2}
\Biggr\},
\end{equation}
where 
\begin{equation}
\begin{split}
N_1^\sigma &=  4m (1 - 2x) q^\sigma - 2 \nt{k} \gamma^\sigma \nt{k}
- 2xy \left(2m^2 + q^2 \right) \gamma^\sigma 
- 2 m^2 \gamma^\sigma \\
N_2^\sigma &=  \nt{k} \gamma^\sigma \nt{k} - x^2 q^2 \gamma^\sigma - 2 m^2 x^2
\gamma^\sigma - m(1-2x) q^\sigma 
- m^2 \gamma^\sigma + m Q^\sigma \\
N_3^\sigma &=  \left(k^2 + m^2 (1 + x^2) + x q^2 (1 + x) \right) 
\gamma^\sigma \\
\Delta_1 &=  m^2 - xy q^2 \\
\Delta_2 &= m^2 y^2 - 2xy q^2. 
\end{split}
\end{equation}
We have used the fact that $Q \cdot q = (p + p') \cdot q = 0$.
The terms containing $q^\sigma$ will vanish after performing 
the $x,y$ integrations. 
This is in accord with the Ward identity.

The only term that will contribute to the 
anomalous magnetic moment is the term in $N_2^\sigma$ involving $Q^\sigma$.
Extracting only this term we have
\begin{equation}
\delta \Gamma^\sigma_{\sf{anom.}} =  
\frac{i 36 g^2}{M^2} \int \frac{d^4 k}{(2 \pi)^4} 
\int_0^1 dx dy \Biggl\{ 
\frac{\delta(x+y-1) \epsilon m Q^\sigma}{\left[k^2 - \Delta_1 \right]^2} 
\Biggr\}.
\end{equation}
Applying dimensional regularization, the $k$ integration yields
\begin{equation}
\delta \Gamma^\sigma_{\sf{anom.}} 
= \frac{- 36 g^2}{(4 \pi)^2 M^2} \int_0^1 dx dy \; \delta(x+y-1) 
\Biggl\{ \epsilon m Q^\sigma \left[\frac{2}{\epsilon} 
- \gamma + \ln \left(\frac{4 \pi \mu^2}{\Delta} \right) \right]
\Biggr\}. 
\end{equation}
Keeping only the zeroth order term in $\epsilon$ we have
\begin{equation}
\label{zeroth}
\delta \Gamma^\sigma_{\sf{anom.}} = 
\frac{- 72 m g^2}{(4 \pi)^2 M^2} Q^\sigma. 
\end{equation}
Using the Gordon identity the magnetic moment part is isolated 
\begin{equation}
\delta \Gamma^\sigma_{\sf{anom.}} = 
\frac{72 i m g^2}{(4 \pi)^2 M^2} \sigma^{\sigma \nu} q_\nu 
\equiv \frac{i \sigma^{\sigma \nu} q_\nu}{2 m} F_2(q^2), 
\end{equation}
providing an explicit form of the structure function,
\begin{equation}
F_2(q^2) =  \frac{144 m^2}{M^2} \frac{g^2}{(4 \pi)^2}. 
\end{equation}
Note that to this, the lowest, order $F_2(q^2)$ is independent of 
$q^2$.
The $g$-factor of the muon is defined by
\begin{equation}
\label{gfactor}
\frac{g_\mu - 2}{2} = F_2(0) = \frac{9 m_{\mu}^2 g^2}{M^2 \pi^2}. 
\end{equation}
This expression will allow us to set a bound on the value of 
the antisymmetric tensor coupling \cite{neville1980} by fitting 
to the experimental discrepancy for the muon $g$-factor.

The standard model prediction differs from experiment 
\cite{hertzog0202024} by $\delta a = 25(16) \times 10^{-10}$. 
Using this difference as an upper bound to the antisymmetric tensor 
contribution we have
\begin{equation}
a_\mu (\mbox{antisym}) \leq 25 \times 10^{-10}
\end{equation}
The muon mass has been measured \cite{review2000} to be
$m_\mu = 105.658$ MeV.  Inserted into (\ref{gfactor}),
this yields
\begin{equation}
\label{newcoupling}
\frac{g^2}{M^2} \leq 2.456 \times 10^{-7} \mbox{ GeV}^{-2}
\simeq 2.5 \times 10^{-7} \mbox{ GeV}^{-2}
\end{equation}
which provides an explicit numerical {\it upper} bound on the 
antisymmetric tensor coupling
\footnote{References \cite{hammond2002,hammond1995b} have used
the expression $\mathcal{L}_{\sf{int}} =  - i \sqrt{\frac{\pi G}{12}} \; \kappa \overline{\psi} \sigma_{\mu \nu \lambda} H^{\mu \nu \lambda} \psi$ as their interaction lagrangian rather than the one that we have used in (\ref{actionkalb})
above. In order to facilitate comparison with their results
it should be noted that our bound on $g^2/M^2$ given in (\ref{newcoupling})
translates into a bound of $\kappa \leq 1.19 \times 10^{16}$.}.
Notice that if $g$ is of order 1, then the mass scale is set at
$M \sim 2$ TeV which is 1)  much smaller than the the Planck scale
and 2) may be near the supersymmetry scale.

This simple coupling makes no distinction between the fermion families
and hence it must have effect on the anomalous magnetic moments of
the electron and the tau.
The electron anomalous magnetic moment is given in \cite{review2000}
as
\begin{equation}
a_e = \frac{g_e - 2}{2} = 1159652187 \pm 4 \times 10^{-12} .
\end{equation}
With our upper bound for the coupling, the Kalb-Ramond contribution
to the electron's magnetic moment is
\begin{equation}
a_e (\mbox{antisym}) \lesssim 0.06 \times 10^{-12}
\end{equation}
which is well within the experimental error.
The experimental value for the tau anomaly is not very precisely 
known at present and the effect of the antisymmetric tensor interaction,
while approximately 300 times larger than in the muon case, is 
expected to be unverifiable in the near future.

\section{Conclusion}

We have found that the interaction between a fermion and an 
antisymmetric tensor field such as the one arising in string 
theory and in Einstein-Cartan gravity can solve the problem
with the muon anomalous magnetic moment without having a
significant affect on the electron anomalous magnetic moment 
if the antisymmetric tensor coupling satisfies the bound 
(\ref{newcoupling}).
 
There are many other possible contributions, beyond the standard model, 
to the muon anomalous magnetic moment. The most favoured candidates 
are supersymmetric partners to the standard model spectrum 
\cite{everett0102145,feng,ferrer,ellis}. If these other particles are 
indeed found then the bound on the Kalb-Ramond--fermion coupling 
would become stronger.

\end{document}